\documentclass{phc-proc4-auth}

\usepackage{graphicx}

\usepackage{amssymb}

\begin{document}

\begin{frontmatter}



\title{Mesoscopic flux jumps in MgB$_2$ films visualized by magneto-optical imaging}


\author{A.V.Bobyl$^{a,b}$, D.V.Shantsev$^{a,b}$, Y.M.Galperin$^{a,b}$}, {\AA}. A. F. Olsen$^a$, 
\author{T.H.Johansen$^{a,\!}$\corauthref{cor}},
\corauth[cor]{Corresponding author. E-mail address:~tomhj@fys.uio.no, Fax:~+47~22856422}
\author{W.N.Kang$^{c}$, S.I.Lee$^{c}$}

\address{$^{a}$ Department of Physics, University of Oslo, P. O. Box 1048
Blindern, 0316 Oslo, Norway\\
$^{b}$A. F. Ioffe Physico-Technical Institute, Polytekhnicheskaya 26,
St.Petersburg 194021, Russia\\ 
$^{c}$National Creative Research Initiative Center for Superconductivity, 
Department of Physics, Pohang University of 
Science and Technology, Pohang 790-784, Republic of Korea
}

\begin{abstract}
We report on the first spatially resolved observation of mesoscopic flux jumps in 
superconducting films. Magneto-optical imaging was used to visualize the 
flux penetration in MgB$_2$ films subjected to a slowly varying perpendicular field. 
Below 10 K, flux jumps with typical size 10-20 microns and regular  
shape are found to occur at random locations along the 
flux front. The total number of vortices participating in one jump is varying 
between 50 and 10000. 
Simultaneously, big dendritic jumps with dimensions comparable to 
the sample size ($10^6$--$10^8$ vortices) are also found in this temperature range. 
We believe that both types of jumps result from thermo-magnetic instability. 

\end{abstract}

\end{frontmatter}

Magnetic flux often penetrates type-II superconductors via discrete jumps 
or avalanches where
a few or many vortices hop at once from one position to another.
Sometimes these jumps are small, with $<100$ vortices taking part in one event, 
and the avalanche size distribution is close to power law \cite{altshuler}.
Sometimes macroscopic flux jumps take place accompanied by abundant 
heat dissipated during the flux motion that gives essential positive feedback 
to the avalanche development \cite{mints}.    
Such thermal flux jumps have been found in MgB$_2$ films where they lead to
a peculiar dendritic and highly-branching patterns of magnetic flux \cite{epl}.
These dendritic flux structures start entering the zero-field-cooled (ZFC) film 
after applying a perpendicular field exceeding some threshold value of 2--10~mT. 
In the present work 
we use magneto-optical (MO) imaging to study flux penetration below this threshold field,
i.e. in the absence of dendritic jumps. 
Surprisingly, here we find flux jumps too, only now on a much smaller scale.

Films of MgB$_2$ were fabricated on Al$_2$O$_3$
substrates using pulsed laser deposition \cite{kang}.
Two samples, 400~nm thick  
with lateral dimensions 5$\times$5~mm$^2$, were studied and showed
similar behaviour. 
The films had the critical temperature of $39$~K, and the critical 
current density of $\sim 10^7$A/cm$^2$.
The flux density distribution was visualized
using MO imaging based on the Faraday effect, 
for a review see Ref.~\cite{jooss}. 

A ZFC film at 4~K was placed in a  perpendicular applied field
slowly increasing in time with rate of $6\mu$T/s.
Shown in Fig.1(a) is the flux density distribution, $B({\bf r},7.1$mT), near the film edge
that coincides with the bottom of the image.
The flux front separating the flux-free interior (black) from
the flux-penetrated region near the edge is highly non-uniform.
By observing the dynamics of flux penetration in the MO microscope, one could immediately
see numerous small jumps taking place along the flux front.
The jumps seem instantaneous, thus their duration is smaller than $0.1$s.
To analyse the jumps quantitatively, we took a series of MO images 
with field interval of $0.01$~mT, and 
subtracted every two subsequent images.
Fig.1(b)

\begin{figure}
\centerline{\includegraphics[width=8.0cm]{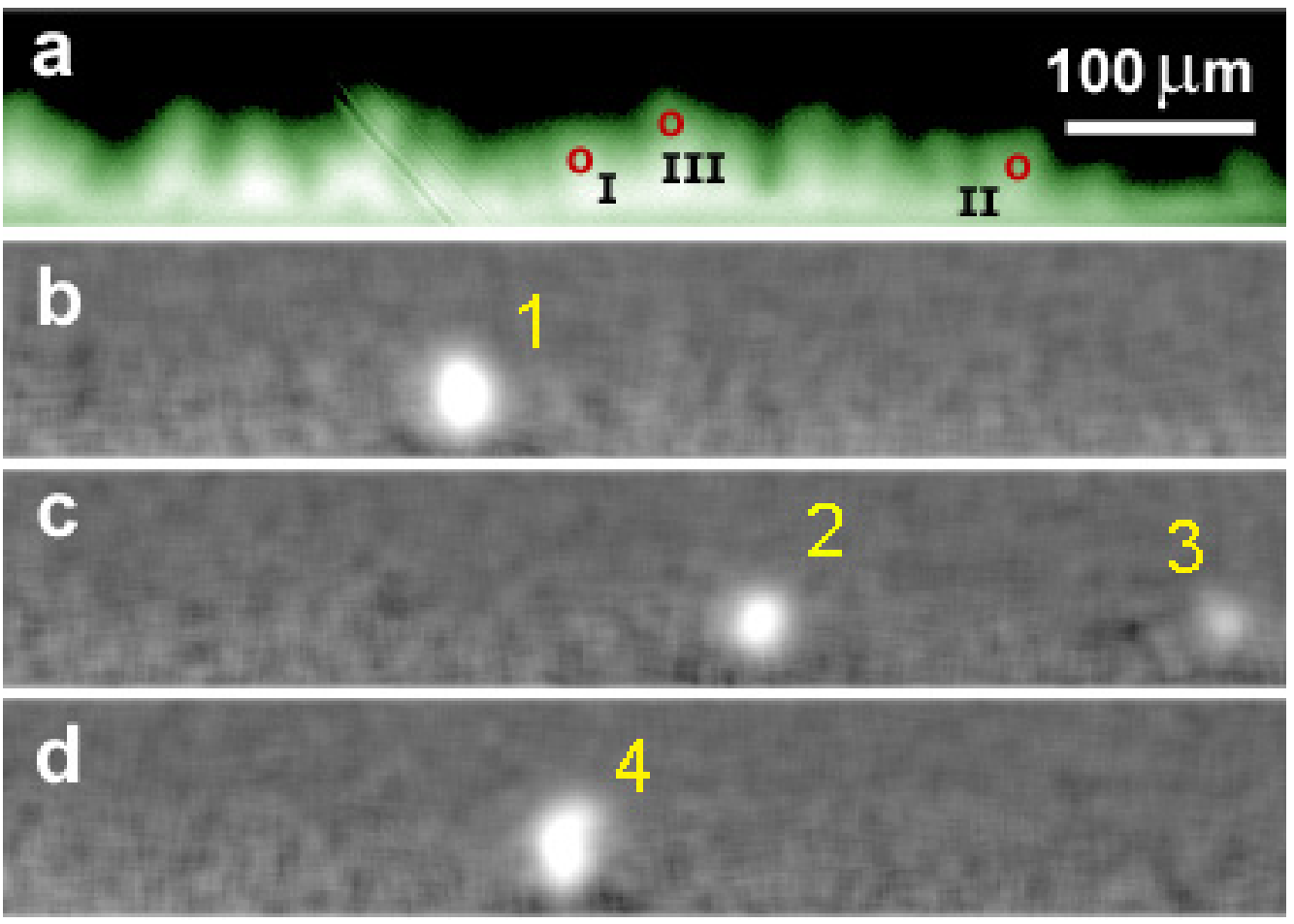}}
\centerline{\includegraphics[width=8.0cm]{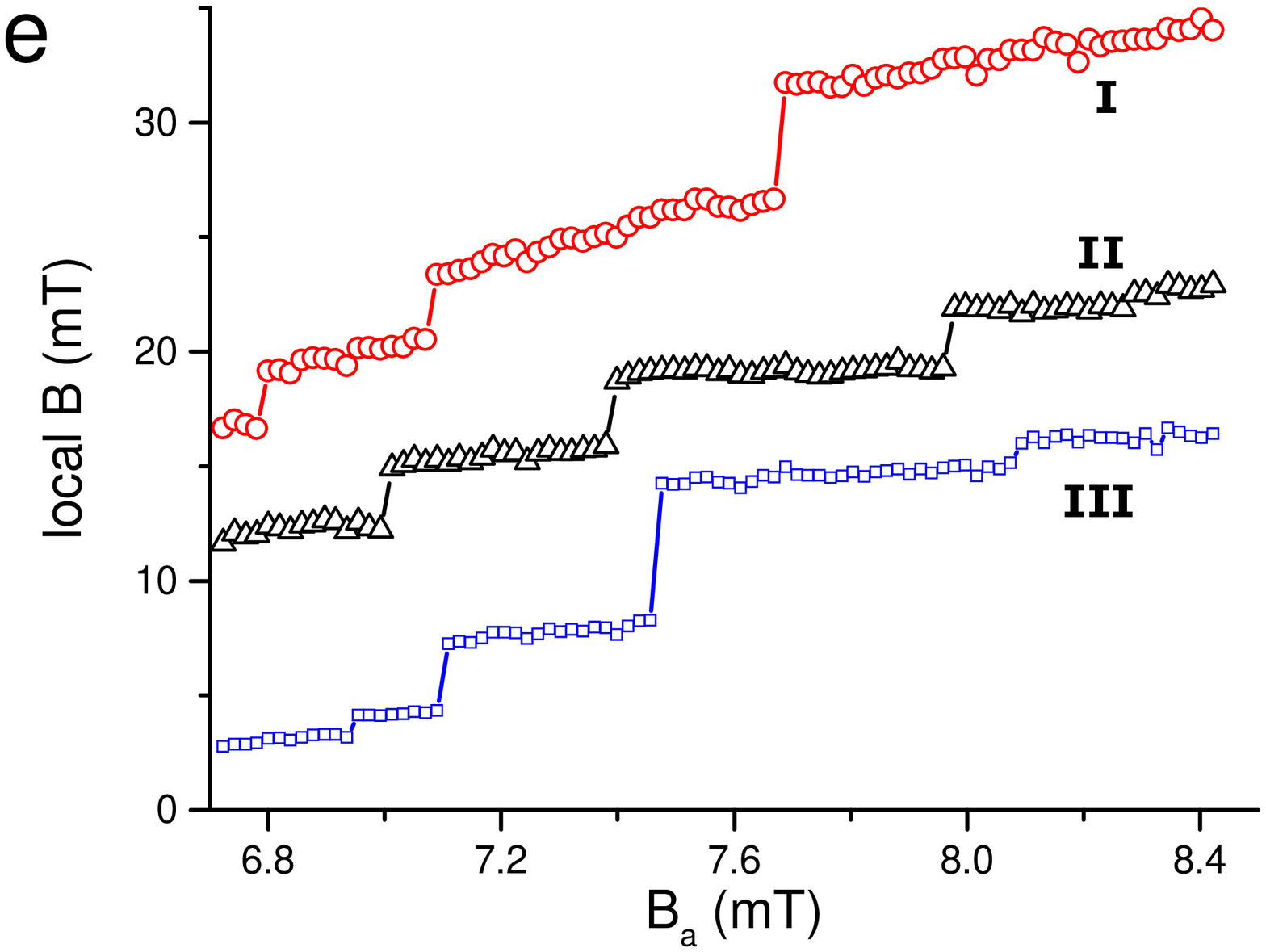}}
\caption{(a): Magneto-optical image of flux distribution $B({\bf r})$
near the edge of MgB$_2$ film in applied field $B_a=7.1$~mT.
(b)-(d): change in $B({\bf r})$ due to increase of $B_a$ by 0.01~mT
as found by subtracting subsequent MO images.
The white spots indicate where localized flux jumps occurred
(the contrast is enhanced). The total amount of incoming flux
found by integrating $B(r)$ over the jump area was  
1900 (1), 1200 (2), 300 (3), and 2100$\Phi_0$ (4).
(e): Local flux density as a function of applied field for three 
regions of $10\times 10 \mu$m$^2$ size 
indicated by circles on the MO image (a). Steps in $B(B_a)$ correspond
to flux jumps.
\label{f}}
\end{figure}

\noindent shows the difference image, $B({\bf r},7.11$mT$)-B({\bf r},7.10$mT), 
where the gray color corresponds to unchanged
flux density, while the white indicates increase in $B$.
One can see that $B$ has increased only in one rather small region.
By integrating the change in 
flux density over this region, we find that the total flux arrived here equals to 
1900$\Phi_0$, $\Phi_0$ is the flux quantum. 
Subsequent difference images (c) and (d) show similar jumps but
in other places along the flux front.    
The first jump with the smallest size resolved by our setup, $50\Phi_0$,
was detected at $B_a=4$~mT. As $B_a$ increases, the average jump size increases too,
the largest jumps being around $10000\Phi_0$. At $B_a=9$mT the first
macroscopic jump of dendritic shape \cite{epl} takes place with size exceeding $10^6\Phi_0$.
The small jumps continue to occur for $B_a>9$mT, but they only slightly affect the 
overall $B({\bf r})$ which is governed now by the dendritic jumps.

To imitate Hall-probe measurements \cite{altshuler} we plot in 
Fig.1(e) local flux density $B$ averaged 
over 10$\times$10~$\mu$m$^2$ area as a function of applied field.
Clear steps dominate the $B(B_a)$ dependence with only little growth of $B$
between them. The steps on different curves corresponding 
to three distinct areas are seemingly uncorrelated.
Moreover, when the whole experiment is repeated, 
the flux distribution $B(r)$ turns out to be well reproduced, 
however the sequence of flux jumps is every time unique.

Two types of jumps different
in the jump size were earlier found in Nb film, and argued to have
different physical mechanisms.\cite{nowak} In our studies
both the small jumps and the macroscopic dendritic jumps 
disappear above the same threshold temperature 10~K. 
This suggests that both types have
the same mechanism -- thermo-magnetic -- and disappear at higher $T$
due to fast increase of the specific heat.       
We explain existence of two types of jumps by an additional positive feedback mechanism
due to strong demagnetization effects in thin films.
When a jump exceeds some critical size, 
bending of current lines around the jump area enhances the local field, and thus 
accelerates the jump development which is then limited only by the sample dimension.

In summary, the flux penetration in MgB$_2$ film below 10~K proceeds via abrupt flux jumps
on all spatial scales. The thermal     
origin of the jumps implies that the apparent critical current density 
is determined here by thermal parameters such as specific heat rather than by 
the pinning strength.

The work was financially supported by NorFa, INTAS 0175wp, FUNMAT/UiO, and the
Research Council of Norway.

\vspace{-0.2cm}




\begin{thebibliography}{00}




\bibitem{altshuler} E. Altshuler, T.H. Johansen, Y. Paltiel, Peng Jin, K.E. Bassler, O. Ramos, G.F. Reiter, E. Zeldov, C.W. Chu
cond-mat/0208266. 

\bibitem{mints} R.G. Mints, A.L. Rakhmanov, Rev. Mod. Phys. {\bf 53}, 551-
592 (1981).

\bibitem{epl} T.H. Johansen et al., 
Europhys. Lett. {\bf 59}, 599 (2002). cond-mat/0104113

\bibitem{kang} W. N. Kang, H. J. Kim, E. M. Choi, C. U. Jung, S. I. Lee, 
Science  {\bf 292}, 1521 (2001). 10.1126/science.1060822.

\bibitem{jooss} Ch. Jooss, J. Albrecht, H. Kuhn, S. Leonhardt and H. Kronmueller,
Rep. Prog. Phys. {\bf 65}, 651 (2002).

\bibitem{nowak} E. R. Nowak, O. W. Taylor, Li Liu, H. M. Jaeger,
T. I. Selinder, Phys. Rev. B {\bf 55}, 11702 (1997).

\end{thebibliography}
\end{document}